\documentstyle[11pt,epsfig]{article} 
\newcommand{\ba}{\begin{array}}
\newcommand{\ea}{\end{array}}
\newcommand{\bd}{\begin{displaymath}}
\newcommand{\ed}{\end{displaymath}}
\newcommand{\be}{\begin{equation}}
\newcommand{\ee}{\end{equation}}
\newcommand{\bea}{\begin{eqnarray}}
\newcommand{\eea}{\end{eqnarray}}

\def\q2 {q^2}

 \def\N10{\widetilde \chi_1^0}
                         \def\C1p{\widetilde \chi_1^+}
                         \def\C1m{\widetilde \chi_1^-}
                         \def\C1pm{\widetilde \chi_1^\pm}
 \def\Ntwo{\widetilde \chi_2^0}
                         \def\Ctwo{\widetilde \chi_2^\pm}
\def\lslep {{\tilde e}_L}
\def\rslep {{\tilde e}_R}
\def\sneu {\tilde \nu}

\def\mET{E_T \hspace{-1.1em}/\;\:}

\def\go{\rightarrow}

\def\lsim{\:\raisebox{-0.5ex}{$\stackrel{\textstyle<}{\sim}$}\:}
\def\gsim{\:\raisebox{-0.5ex}{$\stackrel{\textstyle>}{\sim}$}\:}
\begin{document}
\begin{flushright}
{\large TIFR/TH/01-02 \\ hep-ph/0104217}
\end{flushright}

\begin{center}
{\Large\bf Characteristic Wino Signals in a Linear Collider from
Anomaly Mediated Supersymmetry Breaking}\\[15mm]
{\bf Dilip Kumar Ghosh$^{a,}$\footnote{dghosh@phys.ntu.edu.tw}, 
Anirban Kundu$^{b,}$\footnote{akundu@juphys.ernet.in}, 
Probir Roy$^{c,}$\footnote{probir@tifr.res.in} and 
Sourov Roy$^{c,}$\footnote{sourov@theory.tifr.res.in}}\\[4mm]
{\em $^{a}$ Department of Physics,
National Taiwan University \\
Taipei, TAIWAN}\\[4mm] 
{\em $^{b}$ Department of Physics, 
Jadavpur University, Kolkata - 700 032, INDIA}\\[4mm] 
{\em $^{c}$ Department of Theoretical Physics, 
Tata Institute of Fundamental Research\\
Homi Bhabha Road, Mumbai - 400 005, INDIA} 
\\[7mm]
\end{center}
\begin{abstract}
Though the minimal model of anomaly mediated supersymmetry breaking has
been significantly constrained by recent experimental and theoretical
work, there are still allowed regions of the parameter space for 
moderate to large values of $\tan\beta$. We show that these regions will 
be comprehensively probed in a ${\sqrt s} =1$ TeV $e^+e^-$ linear collider. 
Diagnostic signals to this end are studied by zeroing in on a unique and 
distinct feature of a large class of models in this genre: a neutral 
winolike Lightest Supersymmetric Particle closely degenerate in mass with 
a winolike chargino. The pair production processes $e^+e^- \rightarrow 
{\tilde e}_L^\pm {\tilde e}_L^\mp$, ${\tilde e}_R^\pm {\tilde e}_R^\mp$, 
${\tilde e}_L^\pm {\tilde e}_R^\mp$, ${\tilde \nu} {\bar {\tilde \nu}}$, 
$\widetilde \chi^0_1 \widetilde \chi^0_2$, $\widetilde \chi^0_2 \widetilde 
\chi^0_2$ are all considered at $\sqrt s = 1$ TeV corresponding to the 
proposed TESLA linear collider in two natural categories of mass ordering 
in the sparticle spectra. The signals analysed comprise multiple combinations 
of fast charged leptons (any of which can act as the trigger) plus displaced 
vertices $X_D$ (any of which can be identified by a heavy ionizing track 
terminating in the detector) and/or associated soft pions with characteristic 
momentum distributions.
\end{abstract}

\vskip 1 true cm

\noindent

\newpage
\setcounter{footnote}{0}

\section{Introduction}
The Minimal Supersymmetric Standard Model (MSSM \cite{Rev_SUSY}), with 
softly broken N=1 supersymmetry, is the prime candidate today for 
physics beyond the Standard Model (SM). Suppose the soft supersymmetry 
breaking terms explicitly appearing in the MSSM Lagrangian are regarded 
as low energy remnants of a spontaneous or dynamical breaking of 
supersymmetry at a high scale. We then know that the latter cannot take 
place in the observable sector of MSSM fields; it must occur in a 
{\em hidden or secluded} sector which is generally a singlet under 
SM gauge transformations. Though the precise mechanism of how supersymmetry 
breaking is transmitted to the observable sector of MSSM superfields is 
unknown, signals of supersymmetry in collider experiments sometimes depend 
sensitively on it. There are several different ideas of transmission, such 
as gravitational mediation at the tree level \cite{nilles} and gauge 
mediation \cite{gauge-mediated}, each with its distinct signatures. A very 
interesting recent idea, that is different from the above two, is that
of Anomaly Mediated Supersymmetry Breaking (AMSB) \cite{randall, giudice} 
which is explained below. This scenario has led to a whole class of models 
and their phenomenological implications have been discussed 
[\ref{randall} - \ref{kaplan}], but let us confine ourselves here to the 
minimal version \cite{randall}. The latter has characteristically distinct 
and unique laboratory signatures. These have been explored in some detail 
for hadronic collider processes [\ref{feng} - \ref{tata}]. But there are 
also quite striking signals that such models predict for processes to be 
studied in a high energy $e^+e^-$ (or $\mu^+\mu^-$) linear collider
\cite{dkgprsr}. This paper aims to provide a first detailed study of such 
signals in this type of a machine.

In ordinary gravity-mediated supersymmetry breaking tree level exchanges
with gravitational couplings between the hidden and the observable
sectors transmit supersymmetry breaking from one to the other. If
$\Lambda_{ss}$ is the scale of spontaneous or dynamical breaking of
supersymmetry in the hidden sector, explicit supersymmetry violating
mass parameters of order $M_s \equiv \Lambda^2_{ss} M^{-1}_{pl}$,
$M_{pl}$ being the reduced Planck mass $\sim 2 \times 10^{18}$ GeV, get
generated in the MSSM Lagrangian. However, in general, there are
uncontrolled flavour changing neutral current (FCNC) amplitudes in this
scenario. Though gravity is flavour blind, the supergravity invariance of
the Lagrangian still admits direct interaction terms between the hidden
and the observable sectors, such as

\be
{\cal L}_{eff} \simeq \int d^4 \theta {h \over {M^2_{pl}}} \Sigma^\dagger
\Sigma Q^\dagger Q. 
\label{eq:flavor}
\ee 

\noindent In Eq. (1), $\Sigma$ and $Q$ are generic chiral superfields in 
the hidden and the observable sectors respectively, while $h$ is a
dimensionless coupling of order unity. When supersymmetry is broken in
the hidden sector, say through a nonzero vacuum expectation value
$\langle F_\Sigma \rangle$ of the auxiliary component, masses get induced 
for spin zero sparticles in the observable sector. There is no symmetry to 
keep $h$ diagonal in flavour space. As a result, the squark soft mass
matrix can have significant off-diagonal terms in that space, leading to 
a major conflict with strong experimental upper bounds on FCNC amplitudes 
from $K-{\bar K}$ and $B-{\bar B}$ mixing as well as from unobserved 
$\mu \rightarrow e \gamma$ decay and $\mu \rightarrow e$ conversions in 
nuclei \cite{fcnc}.

\begin{figure}[hbt]
\centerline{\epsfig{file=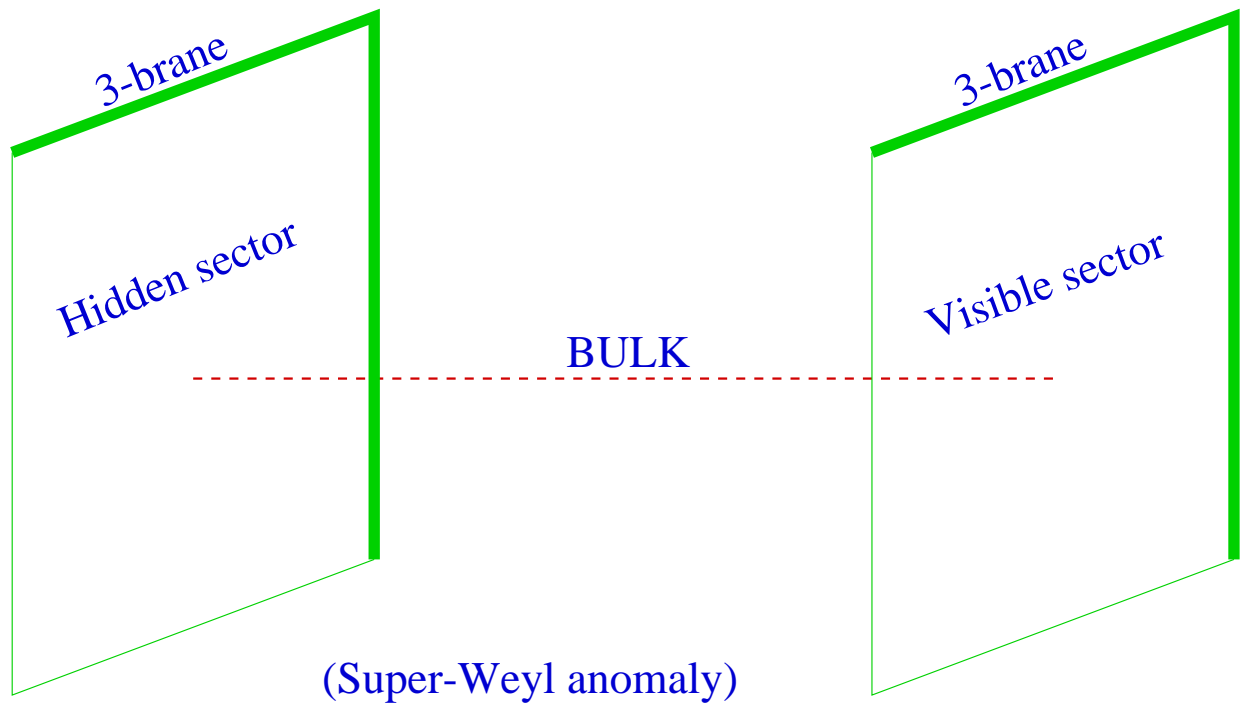,width=\linewidth}}
\caption{\it Supersymmetry breaking across the extra dimension(s).} 
\label{fig:3brane}
\end{figure}

Attempts to solve the above mentioned flavour problem have generally required
special family symmetries \cite{barbieri} in addition to supergravity. 
In the alternative scenario \cite{gauge-mediated}, namely gauge mediated 
supersymmetry breaking, FCNC amplitudes are naturally absent. However, 
models of that category have other difficulties like the CP problem and
the $\mu$ vs $\mu B$ ($\mu$ and $B$ being the higgsino mass and the Higgs
sector supersymmetry violating parameters respectively) problem which
are less severe in gravity-mediated scenarios. It is in this context that 
the recently proposed AMSB scenario has right away become interesting. 
Anomaly mediation is in fact a special case of gravity mediation where 
there are no direct tree level couplings between the superfields of the 
hidden and the observable sectors that convey supersymmetry breaking. This 
is realized, for instance, when the hidden and the observable sector 
superfields are localized on two parallel but distinct 3-branes located in 
a higher dimensional bulk as schematically depicted in Fig. \ref{fig:3brane}. 

Suppose that the two branes are separated by the bulk distance 
$\sim r_c$ where $r_c$ is the compactification radius. Then any tree 
level exchange with bulk fields of mass $m$ ($> r_c^{-1}$) will be 
suppressed by the factor $e^{-m r_c}$. Supergravity fields propagate in 
the bulk, but the supergravity-mediated tree level couplings can now be 
eliminated by a rescaling transformation. The problem can be considered in 
the background of a conformal compensator superfield $\Phi$ whose VEV is 
given \cite{randall} by $\langle \Phi \rangle$ = 1 + $m_{3/2}\theta \theta$, 
$m_{3/2}$ being the gravitino mass. The rescaling transformation is then 
given by $Q\Phi \rightarrow Q$. However, this rescaling symmetry is 
anomalous at the quantum level and the communication of supersymmetry 
breaking from the hidden to the observable sector takes place through the 
loop-generated superconformal anomaly. The latter is topological in origin 
and naturally conserves flavour so that no new FCNC amplitudes are 
introduced from supersymmetry breaking terms. Thus other advantages of 
gravity mediated supersymmetry breaking are retained in the AMSB scenario 
while the flavour problem is solved.   

Let us assume that all explicit soft supersymmetry breaking parameters of 
the MSSM originate from AMSB. Then the masses of the gauginos and of the 
scalars get generated at the same order in the corresponding gauge coupling
strengths. Therefore these masses are not expected to be very different. The 
expressions for the said quantities, in fact, become renormalization group 
invariant with sleptons becoming tachyonic. The latter fact is the most 
severe problem in the AMSB scenario and there are several proposals in 
the literature to solve it. We shall focus on the minimal \cite{randall} 
model of AMSB where the problem of tachyonic sleptons is tackled by adding 
a universal constant term $m^2_0$ to the expressions for squared scalar 
masses, {\em i.e.}, $m^2_0$ contributes equally to the squared masses of 
all scalars present in the theory. One can now solve the tachyonic slepton 
problem through a suitably chosen $m^2_0$ making all squared slepton masses 
positive, the scale invariance of the expressions for scalar masses being 
lost.  The evolution of scalar masses governed by the corresponding 
renormalization group equations (RGE), starting from 
a very high energy scale, must therefore be taken into account. However, 
except for the addition of an extra parameter, this is quite a feasible 
procedure. Recently, a mechanism to generate the minimal AMSB spectrum 
(including the universal additive constant $m^2_0$ in all squared scalar 
masses) has been proposed \cite{kaplan}. In this scheme only the SM matter 
superfields are confined to the 3-brane of the observable sector while the 
gauge superfields live in the bulk along with gravity. The presence of gauge
and gaugino fields in the bulk results in the additional universal
supersymmetry breaking contribution to all scalar masses squared. This
contribution is naturally of the same size as its anomaly-induced
counterparts.   

One can say, on the basis of this last discussion, that the minimal AMSB
model has now been soundly formulated and is worthy of serious
consideration. Indeed, its parameter space has been severely constrained
[\ref{g-22.61} - \ref{g-22.64}] by recent measurements of the $g-2$ of the 
muon \cite{recentg-2} and of the branching fraction for the decay 
$B \rightarrow X_s \gamma$, but some regions of the parameter space 
still survive. This model is characterized by several distinct features 
with important phenomenological consequences: a rather massive gravitino 
($\sim$ a few TeV) and nearly mass-degenerate left and right selectrons and 
smuons while the staus split into two distinct mass eigenstates with the 
${\tilde \tau}_1$ being the lightest charged slepton. 
Most importantly, the model has gaugino masses proportional to the 
$\beta$-functions of the gauge couplings. The latter leads to the existence 
of a neutral near-Wino as the lightest supersymmetric particle (LSP) 
$\N10$ and closely mass-degenerate with it a pair of charged near-Winos as 
the lighter charginos $\C1pm$. A tiny mass difference $\Delta M$ $(<1$ GeV) 
arises between them from loop corrections and a weak gaugino-higgsino mixing 
at the tree level. Because of the small magnitude of $\Delta M$, $\C1pm$, 
if produced in a detector, will be long-lived. Such a chargino then is likely 
to leave \cite{chendreesguni} a displaced vertex $X_D$ and/or a characteristic 
soft pion from the decay $\C1pm \rightarrow \N10 + \pi^\pm$.

The pair production of MSSM sparticles and their decays which are
characteristic of AMSB scenarios, have been considered in various
theoretical studies conducted with respect to experiments in a hadronic
collider [\ref{feng} - \ref{tata}]. But the same can be carried out for
electron and muon colliders too \cite{dkgprsr}. Let us specifically take 
an $e^+e^-$ linear collider at a CM energy $\sqrt{s} = 1$ TeV; the
reason for such a choice will be explained below. We consider 
pair production processes such as $e^+ e^- \rightarrow \lslep^\pm \lslep^\mp$, 
$\rslep^\pm \rslep^\mp$, $\lslep^\pm \rslep^\mp$, $\sneu \bar {\sneu}$, 
$\N10 \Ntwo$, $\Ntwo\Ntwo$. One could also produce smuon or stau pairs and 
look at the signals coming from their cascade decays.  In case of smuons 
the event rates would be reduced typically by a factor of five on account 
of $s$-channel suppression. This is the reason why we have not explicitly 
considered smuon pair production here, though one can look for it in 
principle. A similar argument holds for stau pair production. 
Also, for the sake of simplicity, we choose not to consider taus in the 
final state here because of the lesser experimental efficiency in identifying
them. There are broadly two types of AMSB mass spectra which we 
call Spectrum A and Spectrum B, to be described later. The cascade decay 
chains of the produced sparticles are different for the two spectra. We 
consider both spectra in detail and provide quantitative discussions of 
signals with final states containing multiple combinations of 
$X_D$/soft pions and fast charged leptons in both cases. As mentioned 
earlier, significant constraints [\ref{g-22.61} - \ref{g-22.64}] on the 
minimal AMSB model have been derived recently from measurements of the 
rare deay rate $\Gamma(B \rightarrow X_s \gamma)$ and of the muon anomalous
magnetic moment ${(g-2)}_\mu$. Specifically, negative values of the
Higgsino mass parameter $\mu$ have been effectively ruled out. However,
we shall show that some regions of the AMSB parameter space are
still allowed for moderate to large values of $\tan\beta$ (the ratio of
the two Higgs VEVs). These are precisely the regions that we target in
our studies. Moreover, the required absence of charge and colour
violating global minima suggests \cite{akundu} that charged sleptons in
the minimal AMSB model are rather heavy and beyond the reach of a ${\sqrt
s} = 500$ GeV linear collider, the proposed JLC and NLC machines. It is
interesting to note that the region still allowed by the ${(g-2)}_\mu$
and the $B \rightarrow X_s \gamma$ constraints is almost the one allowed
from more theoretically motivated stability conditions. This is 
why we have chosen a $\sqrt{s}$ value of 1 TeV, to be attained by a linear 
collider such as the proposed TESLA \cite{zerwas}. Charged sleptons of the 
minimal AMSB model are therefore likely to be above (within) the kinematic 
reach of the former (latter) machine. 

There have, of course, been other proposed models to solve the problem
of tachyonic sleptons [\ref{pomarol} - \ref{jackjones}]. In most of
these, however, the sparticle mass spectrum differs from that of the
minimal model both in the gaugino as well as in the scalar sector. Since
we are interested only in a near-Wino LSP scenario, we shall not discuss
the phenomenology associated with those models\footnote{See, for
example, the second paper in \cite{pomarol}.} which lack this feature.
The near-Wino LSP feature is retained, however, in models invoking
additional $D$-term contributions to soft scalar masses
[\ref{katz} - \ref{carena}] keeping the sum of squared scalar masses RG
invariant. This is because the mass spectrum in the gaugino sector
remains the same as in the minimal AMSB model. Thus such schemes will
lead finally to the kind of signals in a high energy $e^+e^-$ linear
collider\footnote{Similar considerations apply to a future $\mu^+ \mu^-$
collider with the difference that the pair production of smuons will be
more copious than that of selectrons.} that we are going to discuss in 
the minimal model. However, the extreme degeneracy of the right and left 
selectrons (smuons) is lost here with the right becoming heavier. This 
feature can change the patterns of the cascade decays of various sparticles 
leading to observable consequences different from those of the minimal 
model. We shall discuss these complications separately. 

The paper is organized as follows. In section 2 we
describe and discuss the mass spectra of the minimal AMSB model and
distinguish between the two cases A and B. In section 3, we enumerate
processes of the pair production of different sleptons and neutralinos, 
as mentioned earlier; we also consider the relevant cascade decay chains 
that lead to our desired signals. The numerical results of these 
computations are presented in section 4 which also contains
discussions of the results. The final section 5 provides a brief summary
of our conclusions.  

\section{Spectra and Couplings of the Minimal AMSB Model}

Scalar masses in the minimal AMSB model are determined via the
RGE equations of the MSSM with appropriate boundary conditions at the 
gauge coupling unification scale $M_G \sim (1.5-2.0) \times 10^{16} 
~{\rm GeV}$. In the present analysis, the evolution of gauge and Yukawa 
couplings as well as that of scalar mass parameters are computed using 
two-loop RGE equations \cite{two-loop}. We have also incorporated the 
unification of gauge couplings at that scale with $\alpha_3(M_Z) 
\approx 0.118$. The magnitude of the higgsino mass parameter $\mu$ gets 
fixed from the requirement of a radiative electroweak symmetry breaking. 
It is computed at the complete one-loop level of the effective potential
\cite{one-loop}, the optimal choice for the 
renormalization scale being $Q = \sqrt{(m_{{\tilde t}_1} m_{{\tilde t}_2})}$, 
$m_{{\tilde t}_1} (m_{{\tilde t}_2})$ being the lighter (heavier) stop 
mass. 
We have also included a supersymmetric QCD correction to the
bottom-quark mass \cite{susyqcd}, which is significant for large
$\tan\beta$. The model has only four parameters: the gravitino mass
$m_{3/2}$, the common scalar mass parameter $m_0$, $\tan\beta$ and
sgn($\mu$). However, we exclude the case $\mu < 0$ since that has now
been ruled out \cite{g-22.61} by the recent ${(g-2)}_\mu$ data. The 
appropriate boundary conditions at the GUT scale for the masses and the 
trilinear couplings are given below. It should be re-emphasized at this 
point that gaugino masses and trilinear scalar couplings can be computed 
from these expressions at any scale once the appropriate values of the 
gauge ($g$) and Yukawa ($h$) couplings at that scale are known. We shall 
ignore Yukawa terms in the superpotential (and also the trilinear scalar 
couplings) pertaining to the first two generations, though not those 
involving the third generation. Sfermions of the heaviest family are 
therefore treated differently from those of the two lighter ones.  

The scale invariant one-loop gaugino mass expressions are  
\be
M_i = b_i {g^2_i \over 16 \pi^2} m_{3/2}.
\label{eq:gaugino}
\ee
where $b_1=33/5$, $b_2=1$, $b_3=-3$. 
To leading order, $M_{1,2,3}$ are independent of $m_0$ and
increase linearly with $m_{3/2}$. Furthermore, at one-loop level
the squared masses for the Higgs and third generation scalars are 

\be
m^2_i = C_i {m^2_{3/2} \over {(16 \pi^2)}^2} + m^2_0, 
\label{eq:scalar}
\ee
where $i \equiv (Q,{\bar U},{\bar D},L,{\bar E},H_u,H_d)$ in standard 
notation, with the $C_i$'s being given as 

\bea
\label{eq:cis}
C_Q &=& -{11 \over 50} g^4_1 - {3 \over 2} g^4_2 + 8 g^4_3 + h_t {\hat
\beta}_{h_t} + h_b {\hat \beta}_{h_b},\nonumber\\
C_{\bar U} &=& -{88 \over 25} g^4_1 + 8 g^4_3 + 2 h_t {\hat \beta}_{h_t}, 
\nonumber\\
C_{\bar D} &=& -{22 \over 25} g^4_1 + 8 g^4_3 + 2 h_b {\hat \beta}_{h_b},
\nonumber\\
C_L &=& -{99 \over 50} g^4_1 -{3 \over 2} g^4_2 + h_\tau {\hat 
\beta}_{h_\tau},\\
C_{\bar E} &=& -{198 \over 25} g^4_1 + 2 h_\tau {\hat \beta}_{h_\tau},
\nonumber\\
C_{H_u} &=& -{99 \over 50} g^4_1 -{3 \over 2} g^4_2 + 3 h_t {\hat 
\beta}_{h_t},
\nonumber\\
C_{H_d} &=& -{99 \over 50} g^4_1 -{3 \over 2} g^4_2 + 3 h_b {\hat
\beta}_{h_b} + h_\tau {\hat \beta}_{h_\tau}.\nonumber
\eea
\noindent In Eqs. (\ref{eq:cis}), $Q$ and $L$ refer to the respective squark 
and slepton $SU(2)$ doublet superfields, while ${\bar U}$, ${\bar D}$ and 
${\bar E}$ stand for the singlet up-squark, down-squark and charged slepton 
superfields respectively and $H_u$ ($H_d$) describes the up-type
(down-type) Higgs superfield. Moreover, the ${\hat \beta}$'s are given by
\bea
\hat \beta = {(16 \pi^2)}^{-1} \beta, 
\eea
where $\beta$ is the usual beta-function. Thus 
\bea
{\hat \beta}_{h_t} &=& h_t \left(-{13 \over 15}g^2_1 -3 g^2_2 - {16 
\over 3} g^2_3 + 6 h^2_t + h^2_b \right),\nonumber\\
{\hat \beta}_{h_b} &=& h_b \left(-{7 \over 15}g^2_1 -3 g^2_2 - {16 
\over 3} g^2_3 + h^2_t + 6 h^2_b + h^2_{\tau} \right),\nonumber\\
{\hat \beta}_{h_\tau} &=& h_\tau \left(-{9 \over 5}g^2_1 -3 g^2_2  
+ 3 h^2_b + 4 h^2_{\tau} \right).\nonumber 
\eea
\noindent Finally, the scale invariant expressions for the third generation 
trilinear $A$ couplings are 
\be
\label{eq:aterm}
A_t = {{\hat \beta}_{h_t} \over h_t} {m_{3/2} \over 16 \pi^2},~A_b =
{{\hat \beta}_{h_b} \over h_b} {m_{3/2} \over 16 \pi^2},~ A_{\tau} 
= {{\hat \beta}_{h_{\tau}} \over h_{\tau}} {m_{3/2} \over 16 \pi^2}.
\ee

\noindent The expressions for sfermion masses and $A$-parameters of the
first two generations can be obtained from the above by simply dropping
the Yukawa couplings.

One of the interesting features of the minimal AMSB model is that the
ratios of the respective $SU(3)$, $SU(2)$ and $U(1)$ gaugino mass
parameters $M_3$, $M_2$ and $M_1$ are given, after
taking into account the next to leading order (NLO) corrections and
the weak scale threshold corrections \cite{wells}, by  
\be
\label{eq:ratio}
|M_1| : |M_2| : |M_3| = 2.8 : 1: 7.1
\ee
(Note that eq.\ (\ref{eq:ratio}), as deduced in \cite{wells}, is true only
for low and moderate values of $\tan\beta$. For large values of $\tan\beta$
where bottom and $\tau$ Yukawa couplings are nonnegligible, the ratio
$|M_1|/|M_2|$ slightly increases \cite{tata}.)
As a result of eq.\ (\ref{eq:ratio}), the lighter chargino $\C1pm$ 
and the lightest neutralino $\N10$ become almost degenerate. The 
small difference in the masses, with the lightest neutralino being the
lightest supersymmetric particle (LSP), comes from the tree-level
gaugino-higgsino mixing as well as from the one-loop corrected chargino and 
the neutralino mass matrix. The small mass-splitting $\Delta M$ can be 
approximately written as 
\bea
\Delta M = \frac{ M_W^4 \tan^2\theta_W}{(M_1 - M_2)\mu^2 } \sin^2 2\beta
\bigg [1+ {\cal O } \left(\frac{M_2}{\mu},\frac{M^2_W}{\mu M_1}
\right)\bigg ] \nonumber
      \\[1.2ex]
+ \frac {\alpha M_2}{\pi\sin^2\theta_W}\bigg
[f\left(\frac{M_W^2}{M_2^2} \right)- \cos^2\theta_W f\left(\frac{M_Z^2}
{M_2^2}\right)\bigg ],
\label{eq:delm}
\eea
with
\be
f(x) \equiv
-\frac{x}{4}+\frac{x^2}{8}\ln(x) +\frac{1}{2}\left(1+\frac{x}{2} \right)
\sqrt{4x-x^2} \bigg[\tan^{-1}\left(\frac{2-x}{\sqrt{4x-x^2}}
\right)-\tan^{-1}\left(\frac{x} {\sqrt{4x-x^2}} \right)\bigg ]\nonumber
\ee
The second term in the RHS of eq.\ (\ref{eq:delm}) is the one-loop
contribution which is dominated by gauge boson loops.

\begin{figure}[hbt]
\vspace{-1.0in}
\centerline{\epsfig{file=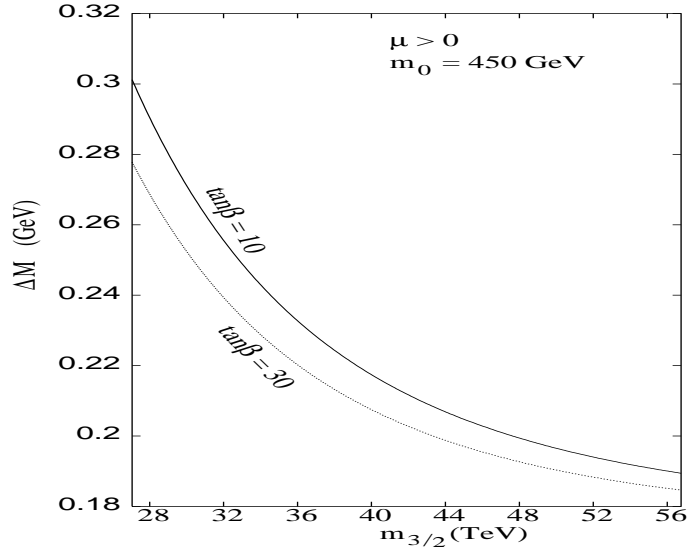,width=10cm}}
\vspace{-1.5in}
\caption{Mass difference $\Delta M$ between the lighter chargino
($\C1pm$) and the LSP ($\N10$) as a function of the gravitino mass
($m_{3/2}$) for $\tan\beta = 10$ (upper curve) and $\tan\beta = 30$
(lower curve), $\mu >0$ and $m_0 = 450$ GeV.}
\label{deltam}
\end{figure}

We have numerically investigated the mass-splitting $\Delta M$ in
various region of the parameter space. Given the LEP lower limit
\cite{charginomass} 
of 86 GeV on the mass of the lighter chargino for 
{\em nearly degenerate} $\C1pm$-$\N10$, it can be 
concluded that the upper limit on $\Delta M$ cannot be much in excess 
of $500~{\rm MeV}$. On the other hand, the condition of radiative electroweak
symmetry breaking in the AMSB scenario gives the ratio $ |\mu/M_2| $ to
be approximately between 3 and 6. For very large $M_2$ (i.e. for very
large $\mu$) the mass difference $\Delta M$ reaches an asymptotic value
of $\approx 165$ MeV. Fig. \ref{deltam} shows the typical variation of
$\Delta M$ as a function of $m_{3/2}$ for a choice of 
$\mu > 0$ and $m_0 = 450$ GeV. The two choices of $\tan\beta$ are
shown in the figure. The curve is cut off on the left by the
lower bound\footnote{The lower bounds on $m_{3/2}$ for two different choices
of $\tan\beta$ are a little different. However, in this figure we have 
taken the same origin for both the curves.} on $m_{3/2}$ implied by the 
experimental constraint \cite{charginomass} on the $\C1pm$ mass ($m_{\C1pm} 
> 86$ GeV) for the nearly degenerate $\C1pm$-$\N10$ case. On the right, it 
gets terminated because otherwise the stau becomes the LSP with a further 
increase\footnote{The value of $m_{3/2}$ for this to happen is higher 
for $\tan\beta =10$ but here we have terminated it at the value 
corresponding to $\tan\beta =30$.} in $m_{3/2}$ for this particular value 
of $m_0$. This is also the reason why we have not been able to show the 
limiting value of $\Delta M \approx 165$ MeV on this plot.

\begin{figure}[ht]
\centerline{\epsfig{file=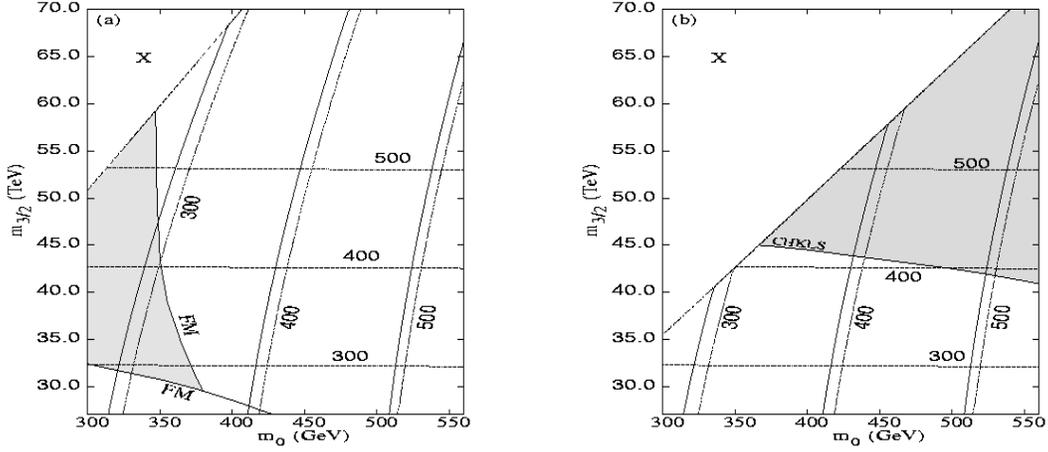,width=11cm}}
\vspace{-2.5in}
\caption{\it Contours of constant charged slepton (continuous) and sneutrino 
(dashed) masses in the $m_0 - m_{3/2}$ plane for (a) $\tan\beta = 10$ and 
$\mu > 0$, (b) $\tan\beta = 30$ and $\mu > 0$. The number adjacent to each 
pair of curved lines corresponds to the sneutrino (slepton) mass for that 
particular contour. The dotted horizontal lines are the contours of constant 
$\Ntwo$ mass. The number adjacent to each horizontal line is the mass of 
$\Ntwo$ for that contour. The boundaries of constraint, labelled by FM and 
CHKLS (the near vertical boundary in (a) from ${(g-2)}_\mu$ and the
near horizontal boundaries in (a) and (b) from $\Gamma (B \rightarrow X_s
\gamma)$), have been taken from Refs. \cite{g-22.61} and \cite{g-22.63}
respectively. The shaded regions are the allowed regions in the two
figures.} 
\label{fig:spectrum1}
\end{figure}

In Fig. \ref{fig:spectrum1}, we show a sample plot of various sparticle 
contours in the $m_0 - m_{3/2}$ plane for two different choices of the other 
parameter, namely $\tan\beta$. The sign of $\mu$ has been taken to be
positive. One of the very striking features of the minimal AMSB model 
is the strong mass degeneracy between the left and right charged sleptons. 
As a result of this, the third and the second 
generation L-R mixing angles become substantial. They can reach the maximal 
limit provided $\tan\beta$ is large. This large L-R mixing of the smuons has 
been shown to have a strong influence on the neutralino loop contribution
(which is a rather small effect as compared to the dominant chargino loop
contribution) to ${(g-2)}_\mu$ \cite{ucdgsr} in the context of the minimal 
AMSB scenario. The corresponding constraint on the parameter space has 
been derived assuming that the supersymmetric contribution to 
${(g-2)}_\mu$ is restricted within a 2$\sigma$ limit of the combined error 
from the Standard Model calculation and from the current uncertainty in 
the Brookhaven measurement [\ref{g-22.61} - \ref{g-22.64}]. A similar 
constraint \cite{g-22.61,g-22.63} from the measured radiative decay rate 
$\Gamma (B \rightarrow X_s \gamma)$ has been invoked.    

The top left corners of both the diagrams in Fig. \ref{fig:spectrum1}, 
marked by $X$, are not allowed. This is due to the twin requirements of
${\tilde \tau}_1$ not being allowed to be the LSP and $m_{{\tilde \tau}_1}$
having to exceed the experimental lower bound of $70$ GeV \cite{stauexp}. 
Any further increase in $m_{3/2}$ for a particular value of $m_0$ is 
disallowed since then the staus become tachyonic. Furthermore, the lower 
limit on $m_{3/2}$ in each figure comes from the experimental constraint
\cite{charginomass} $m_{\C1pm} > 86$ GeV. One can also check that 
the maximum possible value of $m_{3/2}$ for a given 
$m_0$ is a decreasing function of $\tan\beta$ \cite{ucdgsr}. 
Recently, new bounds on minimal AMSB model parameters 
have been proposed from the condition that the electroweak symmetry breaking 
minimum of the scalar potential is the deepest point in the field space 
\cite{akundu}. Selected parameters for our entire numerical calculation
have been chosen so as to be consistent with all these bounds. 

The continuous and dashed curved lines in Fig. \ref{fig:spectrum1} stand for 
contours of constant charged slepton and sneutrino masses respectively
\footnote{We have liberally allowed a certain amount of uncertainty in the 
theoretical lower bounds \cite{akundu} on $m_0$ and $m_{3/2}$ from the 
required absence of color or charge nonconserving vacua in the one-loop 
effective potential on account of two-loop as well as weak threshold 
corrections.}. It is easy to see why the latter are lower than the former. 
The sneutrino mass is linked with the left charged slepton mass through 
the $SU(2)_L$ relation 
\bea
m^2_{\sneu} = m^2_{\tilde L} + {1 \over 2} m^2_Z cos 2\beta.
\eea
Thus, for $\tan\beta > 1$, $m_{\sneu}$ is always smaller than $m_{\tilde
L}$. Since the right and the left sleptons are very highly degenerate,
sneutrinos are always lighter than the right sleptons too. The horizontal 
lines in Fig. \ref{fig:spectrum1} are contours of constant $\Ntwo$ mass. 
It may be noted that no $\N10$ mass contour has entered these figures 
since the LSP $\N10$ is always significantly lighter than $\Ntwo$. 
Depending on various parametric choices, we see that the AMSB sparticle mass 
spectrum can be broadly classified into two natural categories with mass 
ordering as given below (sparticle symbols standing for sparticle masses):

\begin{itemize}
\item Spectrum A: $\N10(\approx \C1pm)
< \tilde \nu < {\tilde e}_R (\approx {\tilde e}_L) < \Ntwo$
\item Spectrum B: $\N10(\approx \C1pm)
< \Ntwo < \tilde \nu < {\tilde e}_R (\approx {\tilde e}_L).$
\end{itemize}
Note that, in AMSB scenarios with a wino LSP, $\Ctwo$ and $\widetilde
\chi^0_{3,4}$ are higgsino-dominated and heavy enough to be neglected in
our signal studies. Also, for large values of $\tan\beta$, there exists
a region where the mass of $\Ntwo$ is between that of the lighter stau
$\tilde\tau_1$ and the first generation sleptons. This variant, which
we call Spectrum B1, is characterized by
\begin{itemize}
\item Spectrum B1: $\N10(\approx \C1pm)
< \tilde\tau_1< \Ntwo < \tilde \nu < {\tilde e}_R (\approx {\tilde
e}_L).$
\end{itemize}
For most of the decay modes, the spectra B and B1 have identical
behaviour, and we do not list them separately. The only exception is the 
decay of $\Ntwo$, which is discussed in Section 3.2.
 
It can be seen from Fig. \ref{fig:spectrum1} that mass-orderings, other 
than A and B above, are allowed only in very restricted regions of the 
$m_{3/2} - m_0$ plane and correspond to the fine tuning of some parameters. 
This is why we do not consider them further. As already mentioned, staus 
and smuons are not considered for production here. That is the reason why 
we have not included them in the two classes of spectrum written above. 
(However, for completeness, one may note that the smuons are almost 
degenerate with the selectrons, while the lighter stau {\em is} the lightest 
slepton for large values of $\tan\beta$). Finally, the regions in the 
parameter space, where the LSP can be either the $\tilde \tau_1$ (the two 
stau mass eigenstates are separated because of the large off-diagonal terms 
in the stau mass matrix) or the sneutrino, are disfavoured on cosmological 
grounds. Such LSPs are also incompatible with the stability of the 
supersymmetric scalar potential \cite{akundu}. Thus, there are really only 
two distinct classes of spectra which are relevant and interesting for our 
present study.  

\section{Production and Decays of Sparticles}
The basic production processes that we consider are 
$e^+e^- \rightarrow {\tilde e}_L^\pm {\tilde e}_L^\mp$, ${\tilde
e}_R^\pm {\tilde e}_R^\mp$, ${\tilde e}_L^\pm {\tilde e}_R^\mp$, 
${\tilde \nu} {\bar {\tilde \nu}}$, $\widetilde \chi^0_1 
\widetilde \chi^0_2$, $\widetilde \chi^0_2 \widetilde \chi^0_2$. We have
not included the process $e^+e^- \rightarrow \C1pm {\widetilde
\chi_1^\mp}$ since that will yield two displaced vertices/soft pions 
in addition to $\mET$; the only way to trigger the reaction would be to 
consider an additional initial-state-radiated (ISR) photon 
\cite{chendreesguni}. However, apart from the rate being lower by a factor 
$\alpha$, such a signal could also come from the light higgsino 
($|\mu| \ll |M_{1,2}|$) scenario. The process $e^+e^-\rightarrow 
\chi_1^+\chi_1^-\gamma$ in the AMSB model has been discussed in Ref. 
\cite{ad-shyama}. Another possible process that we have not 
considered is the heavier $e^+e^- \rightarrow \C1pm \tilde \chi^\pm_2$ 
since the dominant decay mode of $\tilde \chi^\pm_2$ in the AMSB scenario 
is $\tilde \chi^\pm_2 \rightarrow \C1pm h$ and then the branching ratio for 
a final configuration with multilepton + soft pion + $\mET$ is rather small.  

In this section we shall also enumerate all possible decay channels of the 
low-lying sparticles $\C1pm$, $\Ntwo$, $\lslep$, $\rslep$ and $\sneu$ that 
result in one or more leptons plus at least one soft charged pion accompanied
by missing energy. Let us highlight some important characteristics,
selection criteria and conventions that we have used.
\begin{itemize}
\item $\C1pm$ and $\N10$ are almost exclusively winos and $\C1pm$ decays
slowly (with a visibly displaced vertex $X_D$) only\footnote{Henceforth
$\pi$ wil be used to denote $X_D$ and/or charged soft pion.} to $\N10$ 
+ soft $\pi^\pm$.
\item $\Ntwo$ is almost a bino; at least, those decays that require a
nonzero wino component of $\Ntwo$ are neglected, while the higgsino
components are unimportant since we are considering only $e$ or $\mu$ 
in the final state. An exception is made for $\Ntwo$ decay in
spectrum B, and is discussed in section 3.2.
\item Whenever one or more two-body decay channels are kinematically 
allowed, we do not consider possible three- or four-body channels. 
Three-body channels, for instance, are significant only when no two-body 
decay channel exists and are considered only in that situation.
\item Written without any suffix, $\sneu$ and $\nu$ indicate both electron
and muon-type sneutrinos. Similarly, $\ell$ ($\ell_1$, $\ell_2$)
stands for $e$ or $\mu$. For simplicity, we do not consider any $\tau$ 
signals. We use the same notation for both (s)neutrino and anti(s)neutrino; 
they can be discriminated from the charge of the associated (s)lepton.
\end{itemize} 

\subsection{Decay cascade for Spectrum A}

For Spectrum A, the allowed decay channels are as follows:
\begin{enumerate}
\item $\Ntwo \go \nu \sneu,~~~\ell_L^\pm {\tilde \ell}_L^\mp,~~~\ell_R^\pm
{\tilde \ell}_R^\mp$.
\item $\lslep \go e \N10,~~\nu_e \C1pm$.
\item $\rslep \go e \widetilde \chi^{0*}_2 \go e\nu\sneu$. Note that $\rslep$
must have a three-body decay since $\N10$ has a vanishing bino component.
Also, the virtual $\widetilde \chi^{0*}_2$ goes into the $\nu\sneu$ channel 
rather than the $\ell_L^\pm {\tilde \ell}_L^\mp$ channel since 
${\tilde \ell}_L$ is never lighter than $\rslep$ in minimal AMSB. 
(In other AMSB models [\ref{katz} - \ref{carena}] where $\rslep$ is heavier 
than ${\tilde \ell}_L$, this mode is also allowed.) However, the $\tau\tilde 
\tau_1$ channel, though included in the calculation of the branching fractions 
of other channels, is not considered explicitly since we do not look at final
states with $\tau$'s.  
\item $\sneu \go \nu \N10,~~\ell^\mp \C1pm$.  
\end{enumerate}

This immediately predicts the decay cascade for low-lying sparticles.
As already pointed out, $\C1pm$ results in a visibly displaced vertex
$X_D$ and/or a soft charged pion. We club $\N10$, which invariably appears
at the end in sparticle dacays, and $\nu$ under $\mET$. Thus, the end
products of various sleptons and $\Ntwo$ are
\begin{enumerate}
\item $\sneu \go \ell^\pm \pi^\mp \mET$ (it can have a completely invisible
mode $\sneu\go\nu\N10$ and thus can act as a virtual LSP).
\item $\lslep \go e\mET,~\pi\mET$.
\item $\rslep \go e\mET,~e\ell^\pm\pi^\mp \mET$.
\item $\Ntwo \go \ell^\pm \pi^\mp \mET,~\ell^+\ell^- \mET,~ 
\ell_1^+\ell_1^-\ell_2^\pm\pi^\mp \mET$ ($\ell_1,\ell_2 = e,\mu$).
\end{enumerate}

We focus on the following six production processes in an $e^+e^-$
linear collider. Only those final states with one or two soft charged
pions accompanied by {\em at least} one charged lepton and
$\mET$ are enlisted; however, only states with one soft pion, and
dileptons with two soft pions, are discussed in detail in the next
section. Note that some of the possible final states, {\em e.g.}, five 
leptons and a soft pion, has negligibly small cross sections due to very 
small branching ratios in various stages of the cascade.

\subsubsection{Production processes}

\noindent {\bf Process 1}: $e^+e^- \go \sneu\bar{\sneu}$ : Possible
final states here are $\ell^\pm\pi^\mp$ or $\ell^+\ell^-\pi^+\pi^-$.
The leptonic flavour is the flavour of the produced $\sneu$; if we
consider only $\sneu_e$ pair production (which will be more copious
than $\sneu_\mu$ pair production due to the $t$-channel $\lslep$
exchange diagram) the final state will have one or two electrons. 

\noindent {\bf Process 2}: $e^+e^- \go \lslep^+ \lslep^-$ : The only 
possible signal here is $e^\pm\pi^\mp$.

\noindent {\bf Process 3}: $e^+e^- \go \rslep^+ \rslep^-$ : There
can be two possible pionic final states in this case, namely, 
$e^+e^-\ell^\pm\pi^\mp$, and $e^+e^-\ell_1^\pm\ell_2^\mp\pi^+\pi^-$. Note 
that here $\ell_1$ and $\ell_2$ may be different since they come from
the flavour-blind $\Ntwo$ decay. Thus, there are four possible combinations 
of two-pion final states.

\noindent {\bf Process 4}: $e^+e^- \go \rslep^\pm \lslep^\mp$ : This
process takes place only through the $t$-channel exchange of $\Ntwo$. The
corresponding final states are $e^\pm\pi^\mp$, $e^+e^-\ell^\pm\pi^\mp$ and
$e^\pm \ell^\mp \pi^+\pi^-$. 
 
\noindent {\bf Process 5}: $e^+e^- \go \N10 \Ntwo$ : The final state
here is essentially that obtained from the cascade decay of $\Ntwo$. 
We will focus on only two generic channels, namely, 
$\ell^\pm \pi^\mp$ and  $\ell_1^+\ell_1^-\ell_2^\pm\pi^\mp$ 
($\ell_1,\ell_2 = e,\mu$). Depending on the flavours of $\ell_1$ and 
$\ell_2$, there are actually six channels.
 
\noindent {\bf Process 6}: $e^+e^- \go \Ntwo \Ntwo$ : The pair production of
$\Ntwo$ produces a rich variety of possible final states. Single pion
final states include 
$\ell^\pm\pi^\mp$, $\ell_1^\pm\ell_2^+\ell_2^-\pi^\mp$ and $\ell_1^\pm
\ell_2^+\ell_2^-\ell_3^+\ell_3^-\pi^\mp$. Note that every possible
configuration necessarily has an odd number of charged leptons. Depending on
the leptonic flavour, one has twelve channels with single soft pions plus one 
or more charged leptons plus $\mET$. There are three two-pion final states 
with an even number of charged leptons: $\ell_1^\pm\ell_2^\mp\pi^+\pi^-$,
$\ell_1^\pm\ell_2^+\ell_2^-\ell_3^\mp\pi^+\pi^-$ and 
$\ell_1^\pm\ell_2^+\ell_2^-\ell_3^+\ell_3^-\ell_4^\mp\pi^+\pi^-$. These
result in fifteen flavour-specific channels.

\subsection{Decay cascade for Spectrum B}

The allowed decay channels for Spectrum B are as follows:
\begin{enumerate}
\item $\lslep \go e \N10,~~e \Ntwo,~~\nu_e \C1pm$.
\item $\rslep \go e \Ntwo$. Thus $\rslep$ has a more prompt decay in
spectrum B than in spectrum A.
\item $\sneu \go \nu \N10,~~\nu\Ntwo,~~\ell^\mp \C1pm$.  
\item $\Ntwo$ does not have the two-body decay channel to a lepton and
a slepton. Its dominant decay modes are: $\Ntwo\go \N10 h$, $\Ntwo\go
\N10 Z$, $\Ntwo\go \C1pm W^{\mp}$, where $h$ is the lightest CP-even
Higgs scalar. The last two modes occur through the tiny wino and/or
higgsino components of the predominantly binolike $\Ntwo$. Still, they 
dominate over the three-body decay channels mediated by virtual sneutrinos 
or left charged sleptons: 
$\Ntwo \go \nu\sneu^* \go \nu\nu\N10,~\nu\ell^\pm\pi^\mp\N10$;
$\Ntwo\go \ell_L^\pm\tilde{\ell_L^*}^\mp \go \ell^+\ell^-\N10,~
\ell^\pm \nu \pi^\mp\N10$. Since an ${\tilde \ell}_R$ cannot decay into 
$\ell \N10$, any ${\tilde \ell}_R$-mediated decay does not occur. The 
two-body modes are, of course, not kinematically suppressed; the mass limit 
of $\C1pm$ governs the minimum mass splitting of $\Ntwo$ and $\N10$ and 
precludes any such suppression.

For Spectrum B1 (see the last paragraph of Section 2), the two body channel 
$\Ntwo\go\tau\tilde \tau_1$ opens up, and the branching ratios of all the 
above-mentioned two-body channels get suppressed. Since we are not
looking at final states with $\tau$'s, it would be hard to detect a $\Ntwo$ 
in this case.
\end{enumerate}

The decay cascades for sleptons and gauginos are: 
\begin{enumerate}
\item $\Ntwo \go \ell^\pm \pi^\mp \mET,~\ell^+\ell^- \mET$. $\Ntwo$ also
has a virtual LSP mode $\Ntwo\go\nu\nu\N10$. The leptons come from the 
decay of $W$ and $Z$; $h$ decays dominantly into $b\bar b$ and 
$\tau\bar\tau$ which we do not consider here. Note that the three-body 
channels also give the same final states, albeit in a tiny fraction. As 
discussed earlier, the branching ratios for these modes suffer a heavy 
suppression once we look at Spectrum B1.
\item $\sneu \go \ell^\pm \pi^\mp \mET,~\ell^+\ell^- \mET$ (and the virtual 
LSP mode listed for spectrum A).
\item $\lslep \go e\mET,~\pi\mET,~e\ell^+\ell^- \mET,
~e\ell^\pm\pi^\mp \mET$. 
\item $\rslep \go e\mET,~e\ell^+\ell^- \mET,~e\ell^\pm\pi^\mp \mET$.
\end{enumerate}

Next, we enlist the signals for the pair production of charged sleptons and 
gauginos. As stated earlier, our signals consist of one or more charged 
leptons (but no $\tau$) plus one or two soft charged pions, accompanied 
with $\mET$.

\subsubsection{Production processes}

\noindent {\bf Process 1}: $e^+e^- \go \sneu\bar{\sneu}$ : Possible
one pion final states are $\ell^\pm\pi^\mp$ and $\ell_1^+\ell_1^-
\ell_2^\pm\pi^\mp$, where $\ell_1$ and $\ell_2$ can both be $e$ or
$\mu$ irrespective of the sneutrino flavour. Thus, there can be six
flavour-specific final states. The only possible two-pion final state
is $\ell_1^\pm\ell_2^\mp\pi^+\pi^-$, which actually is a combination
of three flavour-specific channels. 

\noindent {\bf Process 2}: $e^+e^- \go \lslep^+ \lslep^-$ : 
The one-pion final states are $\ell^\pm\pi^\mp$, $e^+e^-\ell^\pm\pi^\mp$,
$e^\pm\ell^+\ell^-\pi^\mp$, and $e^+e^-\ell_1^+\ell_1^-\ell_2^\pm\pi^\mp$.
Written in full, they indicate two one-lepton, three three-lepton and four 
five-lepton final states. The two-pion states are $e^\pm\ell^\pm\pi^\mp
\pi^\mp$ (note that in this case we may have a like-sign dilepton signal)
and $e^+e^-\ell_1^\pm\ell_2^\pm\pi^\mp\pi^\mp$ (this will result in a 
four-lepton signal, with three of them of the same sign).

\noindent {\bf Process 3}: $e^+e^- \go \rslep^+ \rslep^-$ : 
Right selectrons can only decay to $\Ntwo$, so the possible one-pion 
signals are $e^+e^-\ell^\pm\pi^\mp$ and $e^+e^-\ell_1^+\ell_1^-\ell_2^\pm
\pi^\mp$. Altogether, they add up to six one-pion final states. The only
possible two-pion state is $e^+e^-\ell_1^\pm\ell_2^\pm\pi^\mp\pi^\mp$.

\noindent {\bf Process 4}: $e^+e^- \go \rslep^\pm \lslep^\mp$ : 
There are four generic one-pion final states for this associated production
process, namely, $e^\pm\pi^\mp$, $e^+e^-\ell^\pm\pi^\mp$, $e^\pm\ell^\pm
\ell^\mp\pi^\mp$ and $e^+e^-\ell_1^+\ell_1^-\ell_2^\pm\pi^\mp$, which result
in eight flavour-specific channels.  The two-pion final states can be
$e^\pm\ell^\pm\pi^\mp \pi^\mp$
and $e^+e^-\ell_1^\pm\ell_2^\pm\pi^\mp\pi^\mp$.

\noindent {\bf Process 5}: $e^+e^- \go \N10 \Ntwo$ : 
There is only one possible channel, {\em viz.} $\ell^\pm\pi^\mp$.

\noindent {\bf Process 6}: $e^+e^- \go \Ntwo \Ntwo$ : 
Possible one-pion channels are $\ell^\pm\pi^\mp$ and $\ell_1^+\ell_1^-
\ell_2^\pm\pi^\mp$, resulting in six channels altogether. 

In the three-lepton channel, two leptons of opposite sign and identical 
flavour should have their invariant mass peaked at $m_Z$. The only
possible two-pion channel is $\ell_1^\pm\ell_2^\pm\pi^\mp\pi^\mp$.
Note that this decay may also give a like-sign dilepton signal.

\begin{table}[htb]
\begin{center}
\begin{tabular}{|c|c|c|} \hline
{\bf Spectrum} & {\bf Signals} & {\bf Parent Channels} \\ \hline
 & $e~\pi$ & $\sneu \bar {\sneu},~~\lslep \lslep,~~\lslep 
\rslep,~~\N10 \Ntwo,~~\Ntwo\Ntwo$ \\
 & $\mu~\pi$ & $\sneu \bar {\sneu},~~\N10 \Ntwo,~~ \Ntwo\Ntwo$ \\
{\bf A} & $e~e~\ell~\pi$ & $\rslep \rslep,~~\lslep \rslep,~~\N10 \Ntwo,~~ 
\Ntwo\Ntwo$ \\
 & $\mu~\mu~\ell~\pi$ & $\N10 \Ntwo,~~
\Ntwo\Ntwo$ \\
 & $\ell_1~\ell_1~\ell_2~\ell_2~\ell_3~\pi$ & $\Ntwo\Ntwo$  
($\ell_{1,2,3}=e,~\mu$) \\ \hline
 & $e~\pi$ & $\sneu \bar {\sneu},~~\lslep \lslep,~~\lslep\rslep,~~
\N10 \Ntwo,~~\Ntwo\Ntwo$ \\
 & $\mu~\pi$ & $\sneu \bar {\sneu},~~\lslep\lslep,~~\N10 \Ntwo,~~ \Ntwo\Ntwo$ \\
{\bf B} & $e~\ell_1~\ell_2~\pi$ & $\rslep \rslep,~~\lslep \rslep,~~
  \lslep \lslep,~~ \sneu \bar {\sneu},~~\Ntwo\Ntwo$ \\
 &                       & ($\ell_{1,2}=e,~\mu$)\\
 & $\mu~\mu~\mu~\pi$ & $\Ntwo\Ntwo,~~ \sneu \bar {\sneu}$ \\
 & $e~e~\ell_1~\ell_1~\ell_2~\pi$ & $\lslep \lslep,~~\rslep \rslep,
~~\lslep \rslep$ ($\ell_{1,2}=e,~\mu$)\\ \hline
%
%
%

\end{tabular}
\end{center}
\caption{Possible one or multilepton signal with one soft pion.
All possible combinations of leptonic flavours are to be taken into
account where the flavour is not shown explicitly.} 
\end{table} 

A list of all possible final states and their parent sparticles, as
discussed above, for both Spectra A and B, is given in Table 1 for one pion 
channels. A similar list is given in Table 2 for two pion channels.  In the 
next section, we discuss some of the one-pion signals and the dilepton plus 
dipion signal, in detail, but let us note a few key features right at this 
point.

\begin{itemize}
\item One can sometimes have the same signal for Spectrum A or B; however, 
their sources are different. This means that the production cross-section 
and different distributions will also vary from one spectrum to the other; 
this may help discriminate between them. A useful option may be to use one 
polarized beam when some of the channels would be altogether ruled out. 
\item One must have an odd (even) number of charged leptons produced in 
conjunction with one (two) soft pion(s) in order to maintain charge 
neutrality. However, one can have a maximum of five leptons in the one-pion 
channel for both Spectra A and B. On the other hand, for two-pion channels, 
Spectrum A allows at most six charged leptons, while Spectrum B allows only 
four. The signal cross sections of the multilepton channels (with lepton 
number $\geq$ 4) may, however, be unobservably small with the presently 
designed luminosity of a $\sqrt s = 1$ TeV linear collider. 
\item Three charged lepton plus one soft pion ($3\ell 1\pi$) signals are 
interesting in their own right. Consider the $3\mu 1\pi$ signal. For 
Spectrum B, two opposite sign muons must have their invariant mass peaked at 
$m_Z$, while no such compulsion exists for Spectrum A. This can serve as a 
useful discriminator between these two options.
\item For signals with more than three charged leptons, one must have 
{\em at least} two electrons for Spectrum B, while all of them can be muons 
for Spectrum A. This can be directly traced to the fact that for Spectrum B 
a selectron pair is the parent whereas a $\Ntwo$ pair generates the 
multilepton signal in Spectrum A.
\item For two-pion and two charged lepton final states, one may have more 
like-sign dileptons for Spectrum B than for Spectrum A. For one-pion states, 
one may get a stronger same-flavour like-sign dilepton signal for Spectrum B. 
The reasons are twofold: (a) Spectrum B gives $\Ntwo$ pairs more frequently at
intermediate stages of the cascade and (b) $\Ntwo$ in Spectrum A is expected
to be heavier than in Spectrum B. In fact, for Spectrum A $\Ntwo$ may even
be outside the energy reach of the collider, in which case there will not be 
any like sign dilepton events in the final state. 
\item From theoretical considerations of charge and colour conservation,  
sleptons are expected \cite{akundu} to be somewhat heavy in the minimal 
AMSB model. In fact, the lower bounds on their masses depend on the 
chargino mass. 
\end{itemize} 

\begin{table}[htb]
\begin{center}
\begin{tabular}{|c|c|c|} \hline
{\bf Spectrum} & {\bf Signals} & {\bf Parent Channels} \\ \hline

 & $e~\ell~\pi~\pi$ & $\sneu \bar {\sneu},~~\lslep \rslep, ~~\Ntwo\Ntwo$ \\
 & $\mu~\mu~\pi~\pi$ & $\sneu \bar {\sneu},~~\Ntwo\Ntwo$ \\
{\bf A} & $e~e~\ell_1~\ell_2~\pi~\pi$ & $\rslep \rslep,
 ~~\Ntwo\Ntwo$ ($\ell_{1,2}= e,~\mu$)\\
 & $\mu~\mu~\mu~\ell~\pi~\pi$ & $\Ntwo\Ntwo$ \\
 & n $e$, (6-n) $\mu$, $\pi~\pi$ & $\Ntwo\Ntwo$\\
 & ($0\leq {\rm n} \leq 6$) & \\ \hline
%
%
%
%
%
%
 & $e~\ell~\pi~\pi$ & $\sneu \bar {\sneu},~~\lslep \lslep,
~~\lslep\rslep,~~\Ntwo\Ntwo$ \\
{\bf B} & $\mu~\mu~\pi~\pi$ & $\sneu \bar {\sneu},~~\Ntwo\Ntwo$\\
 & $e~e~\ell_1~\ell_2~\pi~\pi$ & $\rslep \rslep,~~\lslep \lslep,
~~\rslep \lslep$ ($\ell_{1,2}=e,~\mu$)\\ \hline
%
%
\end{tabular}
\end{center}
\caption{The same as in Table 1, with two soft pions.}
\end{table}

Before we conclude this section, let us just mention how the 
decay products change identity when we deviate from the assumptions of 
minimal AMSB and make $\rslep$ sufficiently heavier than $\lslep$. First, 
note that the decay $\rslep \go \lslep$ plus a virtual higgsino-type 
neutralino is suppressed both due to its mass and the selectron coupling 
of the latter. It need not be considered. We can then discuss the three 
following scenarios:
\begin{itemize}
\item Spectrum A$'$: 
      $\N10(\approx \C1pm)< \tilde \nu< \lslep< \rslep< \Ntwo$
\item Spectrum A$''$: 
      $\N10(\approx \C1pm)< \tilde \nu< \lslep< \Ntwo< \rslep$
\item Spectrum B$'$: 
      $\N10(\approx \C1pm)< \Ntwo< \tilde \nu < \lslep < \rslep$
\end{itemize}

\noindent In spectrum A$'$, only $\rslep$ acquires a new decay channel, 
{\em viz.}, $\rslep \go e \ell^+\ell^- \mET$. This is because the virtual 
$\widetilde \chi^{0*}_2$ can now decay into a left charged slepton, which 
is lighter than the corresponding right charged slepton. Thus, one may have 
a $5\ell+\pi$ signal from the pair production of $\rslep$. 
In spectrum A$''$, $\rslep$ has the two-body decay to $\rslep \go e\Ntwo$. 
Apart from the fact that the lifetime of $\rslep$ is significantly shorter 
in this model as compared to spectrum A, the final products are identical 
(assuming that $\rslep$, the heaviest low-lying charged slepton, is within 
the kinematic reach of the machine). The decay pattern of spectrum B$'$ is 
identical to that of spectrum B since the pattern for the latter does not 
depend upon the degeneracy of $\lslep$ and $\rslep$.

\section{Numerical results and discussions} 
Cross sections for the production of various two-sparticle combinations 
have been calculated at an $e^+e^-$ CM energy of 1 TeV for two values 
of $\tan\beta$, namely, 10 and 30, for $\mu > 0$. Here, we would like to 
point out that in our parton level Monte Carlo simulation we have not 
taken into account the ISR and beamstrahlung effects. The said cross
sections, computed under this condition, were multiplied by the proper 
branching fractions of the corresponding decay channels to get the final 
states described in Tables 4 and 5. Coming to numbers, the selection cuts 
that have been used on the decays products are as follows: 

\begin{itemize}
\item The transverse momentum of the lepton(s) $p^\ell_T > 5$ GeV.   
\item The pseudorapidities of the lepton(s) and of the pion $|\eta| < 2.5$.
\item The electron-pion isolation variable $\Delta R = \sqrt{{(\Delta \eta)}^2 
+ {(\Delta \phi)}^2} > 0.4$. 
\item The missing transverse energy $\mET > 20$ GeV. 
\item The transverse momentum of the soft pion $p^\pi_T > 200$ MeV for a 
detectable pion.
\end{itemize}

The kinematic distributions of the final state particles for the
$e^\pm + \pi^\mp + \mET$ signal have been studied for the following 
sample point in the AMSB parameter space corresponding to Spectrum A: 
$m_{3/2}=44$~TeV, $\tan\beta=30$, $\mu > 0$ and $m_0 = 410$~GeV. For
these values of AMSB input parameters, $\Delta M = 198$ MeV. In order to
obtain the total distribution of some kinematic variable $X$ one now
needs to add the contributions from all the individual channels as 
mentioned earlier. The total normalized distribution can be expressed as:

\bea
\label{dist}
\frac{1}{\sigma}\frac{d\sigma}{dX} = \sum^n_{i =1}
\frac{1}{\sigma_i}\frac{d\sigma_i}{dX}
\eea

\begin{table}[ht]
\begin{center}
\begin{tabular}{|c|c|c|c|c|} \hline
Spectrum & Parameter Set & $m_0$ (GeV) & $m_{3/2}$ (TeV) & $\tan\beta$
\\ \hline 
 & (a) & 340 & 44 &  10 \\ 
 & (b) & 350 & 42 &  10 \\ 
 & (c) & 360 & 39 &  10 \\ 
{\bf A} & (d) & 380 & 46 &  30 \\ 
 & (e) & 410 & 44 &  30 \\ 
 & (f) & 450 & 47 &  30 \\ \hline
 & (a) & 330 & 32 & 10 \\
 & (b) & 350 & 33 & 10 \\
{\bf B} & (c) & 360 & 34 & 10 \\
 & (d) & 460 & 44 & 30 \\
 & (e) & 475 & 44 & 30 \\
 & (f) & 510 & 47 & 30  \\ \hline
\end{tabular}
\end{center}
\caption{Selected parameter points with $\mu> 0 $ for computed cross
sections.}
\end{table}

In Eq. \ref{dist}, $X$ can be the transverse momentum $p^e_T$ of the
electron, the decay length $L$ of the chargino or the soft pion
transverse momentum $p^\pi_T$ and $i$ runs from 1 to 6, covering all
processes which give rise to this distribution. The electron $p_T$ 
distribution corresponding to the signal $e^\pm \pi^\mp \mET$, for
spectrum A, is shown in Fig. 4(a) and demonstrates a very wide distribution. 
This highly energetic electron can be used to trigger such events. After 
that one can look for the heavily ionizing charged track by the 
chargino ending in a soft pion in the Silicon Vertex Detector (SVD) 
located very close to the beam pipe. For this purpose, the knowledge of
the chargino decay length is very important. The probability that the 
chargino decays before travelling a distance $\lambda $ is given by 
$P(\lambda) = 1 - \exp(-\lambda/L)$, where $L = c\tau (\beta\gamma )$ is 
the average decay length of the chargino. From this, one can generate 
the actual decay length distribution of the chargino as 
$ \lambda = - L \ln [1 - P(\lambda)]$, where $P(\lambda)$ is generated 
by a random number between $0$ and $1$. In Fig 4(b), we have displayed the 
chargino decay length distribution with the selection cuts mentioned
above. Though most of the events in this figure cluster at lower decay
lengths for which the $\C1pm$ decay would be so prompt as to make the
charged track invisible (in this case the end product soft $\pi$ 
will still be detectable), a substantial number of events do have 
reasonably large decay lengths for which $X_D$ may be visible. 
In case the chargino track is not seen, our signal can still be observed by 
looking at the soft pion impact parameter $b_{\pi}$ which is determined 
primarily by the chargino decay length and $\Delta M$. It turns out 
that, for our chosen point in the AMSB parameter space, the soft pion 
impact parameter is always significantly above the value where it can be 
resolved ($b_{res}\sim 0.1$ cm). Hence, the prospects of resolving the impact 
parameter of the soft pion are quite high. The normalized $p_T$ distribution 
of the soft pion has been shown in Fig 4(c). Most of the events peak at the 
lower $p^\pi_T < 200 $ MeV, which can be understood from the value of
$\Delta M$ for this point in the parameter space. The imposition of 
selection cuts on the soft pion kills the peak of the distribution.
Nevertheless, we are left with a substantial number of events, which can 
be observed. The distribution looks quite similar for Spectrum B and is not 
shown here. 
\begin{table}[ht]
\begin{center}
\begin{tabular}{|c|c|c|c|c|c|c|c|c|} \hline
Signal & PS & \multicolumn{7}{|c|}{Cross Sections (fb)} \\
\cline{3-9}
 & & $ \sneu \bar {\sneu} $ & $ \lslep {\bar {\tilde e}}_L $  & $\rslep 
{\bar {\tilde e}}_R$  & $\lslep {\bar {\tilde e}}_R + \rslep {\bar
{\tilde e}}_L$ & $\N10 \bar {\Ntwo}$ & $\Ntwo \bar {\Ntwo}$  & $ \rm{Total}$\\
\hline
    & $a$               &29.04 &46.7 &- &0.014 &3.4&0.655 &  79.80 \\ 
\cline{2-9}
    & $b$               &29.51 &45.09 &- &0.016 &3.5 &0.733  &  78. 85 \\ 
\cline{2-9}
$ e \pi + \mET $ & $c$  &30.82 &44.44 &- &0.023 &3.16 &0.616  &  79. 05 \\ 
\cline{2-9}
    & $d$               &21.6 &31.63 &- &2.88 $\times 10^{-5}$ &1.86& 0.171& 
55.26    \\ 
\cline{2-9}
    & $e$               &18.91 &24.33 &- &2.57 $\times 10^{-5}$ &0.98 &0.05 & 
44.27 \\ 
\cline{2-9}
    & $f$               &12.36&13.43 &- &2.65 $\times 10^{-5}$ &0.53& 0.01 & 
26.33\\ 
\cline{2-9}
\hline
\hline
               & $a$ &- &- &1.36 $\times 10^{-4}$ &0.011&1.77 & 0.238 &2.01 \\ 
\cline{2-9}
               & $b$ &- &- &3.65 $\times 10^{-4}$ &0.012 &1.60&0.254 &1.86 \\
\cline{2-9}
$ e e \mu \pi + \mET $ & $c$ &- &- &0.00 &0.018 &1.01 &0.166 & 1.19 \\ 
\cline{2-9}
               & $d$ &- &- &0.00 &2.37 $\times 10^{-5}$ &0.014 & 0.035 & 0.049 
\\ \cline{2-9}
               & $e$ &- &- &0.00 &2.57 $\times 10^{-5}$ &0.013 & 0.008 & 0.021 
\\ \cline{2-9}
               & $f$ &- &- &0.00 &2.65 $\times 10^{-5}$ &0.007 & 0.001&0.008 \\ \cline{2-9}
\hline
\hline
               & $a$ &35.09 &- &-&0.0073 &-&0.070 & 35.16       \\ \cline{2-9}
               & $b$ &36.22 &- &-&0.0085 &-&0.088& 36.31        \\ \cline{2-9}
$ e e \pi \pi + \mET $ & $c$ &39.56 &- &-&0.013&-&0.079&39.65      \\ 
\cline{2-9}
               & $d$ &23.72 &- &-&1.41$\time 10^{-5}$ &-&0.016&23.73   \\ 
\cline{2-9}
               & $e$ &21.36 &- &-&2.57$\times 10^{-5}$&-&0.0059& 21.36  \\ 
\cline{2-9}
               & $f$ &14.48 &- &-&1.272 $\times 10^{-5}$&-&0.0031&  14.48\\ 
\cline{2-9}
\hline

\end{tabular}
\normalsize
\end{center}
\caption{Some selected signals in Spectrum A for sample choices of
parameters in Table 3. The contributions from different sources are also 
shown in the Table. Cross sections less than $10^{-4}$ fb are not added 
to the total cross section. Here, PS stands for Parameter Set.}
\end{table}

For reasons of space and practicality, we shall display numerical
results for only a selected subset of the final states listed in section
3 - mainly to get an idea of signal strengths. Specifically, let us
choose the final states $e~\pi~\mET$, $e~e~\mu~\pi~\mET$, 
$e~e~\pi~\pi~\mET$ for both Spectrum A and Spectrum B. 
In Tables 4 and 5, numbers for the cross sections in the three channels
mentioned above are displayed for the spectra A and B respectively
with the AMSB parameter points as selected in Table 3. The individual
processes have widely different contributions to these channels because of the
fact that their individual production cross sections and branching ratios 
in the cascade decays are highly parameter dependent. For example, for the 
AMSB input $(a)$ of spectrum B,

\begin{table}[ht]
\begin{center}
\begin{tabular}{|c|c|c|c|c|c|c|c|c|} \hline
Signal & PS & \multicolumn{7}{|c|}{Cross Sections (fb)} \\
\cline{3-9}
 & & $ \sneu \bar {\sneu} $ & $ \lslep {\bar {\tilde e}}_L $  & $\rslep 
{\bar {\tilde e}}_R$  & $\lslep {\bar {\tilde e}}_R + \rslep {\bar
{\tilde e}}_L$ & $\N10 \bar 
{\Ntwo}$ & $\Ntwo \bar {\Ntwo}$  & $ \rm{Total}$\\
\hline
    & $a $&44.12 &63.87 &- &0.068 & 0.328&0.029 & 108.41 \\ 
\cline{2-9}
    & $b$ &39.03&54.34 &-&0.0759 &0.333 &0.030 & 93.80\\ 
\cline{2-9}
    & $c$&36.07 &49.30 &-&0.0711 &0.317& 0.027& 85.78\\ 
\cline{2-9}
$ e \pi + \mET $ & $c$  &11.86 &11.40 &- &0.011 &0.041&0.0005 &23.31 \\ 
\cline{2-9}
    & $d$   & 9.46 & 7.90 &- &0.013 &0.049 &0.0008 & 66.37\\ 
\cline{2-9}
    & $e$   &3.98 &1.80 &- &0.0080 &0.040& 0.0004& 5.82 \\ 
\hline
\hline
               & $a$ &6.8$\times 10^{-7}$ &0.004 &0.037&0.631&-&0.0085& 0.68 \\ \cline{2-9}
               & $b$ &8$\times 10^{-5}$ &0.013 &0.051&0.738&-&0.009&  0.81 \\ 
\cline{2-9}
               & $c$ & 6.6 $\times 10^{-5}$ &0.011 &0.044&0.709&-&0.008& 0.77\\ \cline{2-9}
$ e e \mu \pi + \mET $ & $d$ &1.9 $\times 10^{-5}$ &1.9 $\times 10^{-4}$ &5.5 
$\times 10^{-4} $ &0.094&-&1.6 $\times 10^{-4}$&  0.094\\ \cline{2-9}
               & $e$ &1.5 $\times 10^{-5}$ &6.1 $\times 10^{-4}$ &7.0 
$\times 10^{-4}$ &0.108 &-&2.5 $\times 10^{-4}$& 0.10  \\ \cline{2-9}
               & $f$ &7.63 $\times 10^{-6}$ &1.5 $\times 10^{-4}$ &1.5 
$\times 10^{-4}$&0.066&-&1.2 $\times 10^{-4}$& 0.066 \\ \cline{2-9}
\hline
\hline
               & $a$ &62.63 &0.007 &-&1.01&-&0.165& 63.81 \\ \cline{2-9}
               & $b$ &54.81 &0.021 &-&1.16&-&0.172& 56.16 \\ \cline{2-9}
               & $c$ &49.9 &0.017 &-&1.09&-&0.155 & 51.16 \\ \cline{2-9}
$ e e \pi \pi + \mET $ & $d$ &13.24 &2.4 $\times 10^{-4}$ &-&0.113&-&0.002& 
13.35\\ \cline{2-9}
               & $e$ &10.6 &7.5 $\times 10^{-4}$ &-&0.131&-&0.003& 10.73  \\ 
\cline{2-9}
               & $f$ &4.16 &1.75 $\times 10^{-4}$ &-&0.075&-&0.001&  4.23\\ 
\cline{2-9}
\hline

\end{tabular}
\normalsize
\end{center}
\caption{Some selected signals in Spectrum B for sample choices of
parameters in Table 3. The contributions from different sources are also 
shown in the Table. Cross sections less that $10^{-4}$ are not added to the 
total cross section. Here, PS stands for Parameter Set.}
\end{table}

\begin{itemize}
\item
$ BR.(\tilde\nu_e \rightarrow e^\mp + \C1pm) \sim 79\% $
\item
$ BR.(\tilde\nu_e \rightarrow \nu + \N10) \sim 20\% $
\item
$ BR.(\tilde\nu_e \rightarrow \nu + \Ntwo) \sim 2.9\times 10^{-4}\% $
\item
$ BR.(\Ntwo \rightarrow \N10 + h^0) \sim 10.8\% $
\item
$ BR.(\Ntwo \rightarrow \C1pm + W^\mp) \sim 77.3\% $
\item
$ BR.(\Ntwo \rightarrow \tau + \tilde \tau_1) \sim 10.2\% $
\item
$ BR.(\Ntwo \rightarrow  \N10 + Z^0) \sim 1.7\% $
\item
$ BR.(\tilde e_L \rightarrow  \N10 + e) \sim 32.5\% $
\item
$ BR.(\tilde e_L \rightarrow  \C1pm + \nu) \sim 67.4\% $
\item
$ BR.(\tilde e_R \rightarrow  \Ntwo + e) \sim 98\% $
\item
$ BR.(\C1pm \rightarrow  \pi^\pm + \N10) \sim 98\% $
\end{itemize}
All these very different branching ratios play a crucial role in
determining the final number. It turns out that  
BR.($\Ntwo \rightarrow \tau + \tilde \tau_1$) increases with 
$\tan\beta$, reducing the signal cross section at larger $\tan\beta$,
since we have not observing $\tau$'s in the final state. Apart from this
complicated dependence of the different branching fractions, the signal 
cross sections tend to decrease with increasing $m_0,$ and $m_{3/2}$, simply
because of phase space suppression. In the worst cases of the signal 
cross sections, assuming an integrated luminosity of
$500~{\rm fb^{-1}}$, one would expect 13165, 4 and 7240 
signal events in the $e\pi+\mET$, $ ee\mu \pi+\mET $ and 
$ ee\pi \pi + \mET$ channels respectively from Spectrum A, while  
2910, 33,and 2115 signal events are predicted from spectrum B for the same
final state configurations. 

\begin{figure}[hbt]
\vspace{-.01in}
\centerline{\epsfig{file=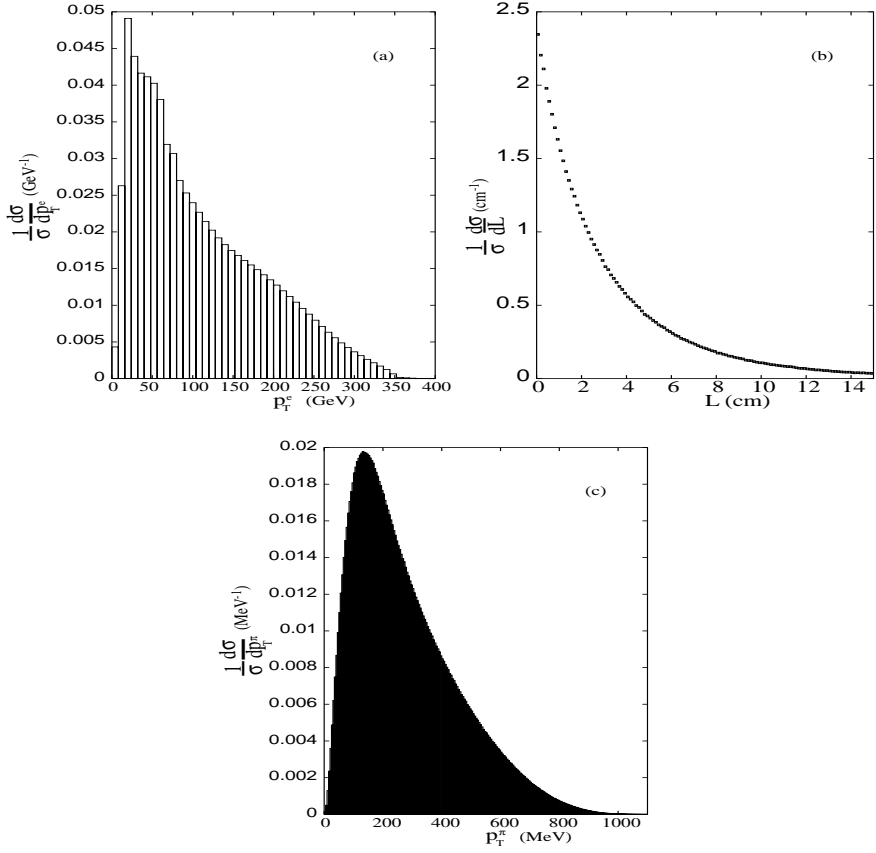,width=8cm}}
\vspace{-.1in}
\caption{Normalized kinematic distributions of decay products:
$(a)$ $p_T$ of charged lepton,
$(b)$ decay length of the lighter chargino,
and $(c)$ $p_T$ of the charged pion arising from
$e^\pm + \pi^\mp + \mET$ signal for spectrum A.
The AMSB input parameters are $m_{3/2}=44$~TeV, $\tan\beta = 30$,
$\mu > 0$ and $m_0 = 410$ GeV.}
\label{kin_dist}
\end{figure}

An alternative MSSM scenario of nearly degenerate $\tilde \chi^0_1$ and
$\tilde \chi^\pm_1$ (and $\tilde \chi^0_2$ as well) can arise
\cite{chendreesguni}
when $|\mu| \ll |M_{1,2}|$. In such a case a mass-difference $\Delta M
(\tilde \chi^\pm_1 - \tilde \chi^0_1) \lsim 1$~GeV can be obtained 
by setting \cite{chendreesguni}
$|M_{1,2}| \gsim 5$~TeV and $|\mu| \gsim {M_Z}/2$. Though this is a
rather unnatural scenario and quite difficult to obtain in a 
phenomenologically viable model, we can ask whether our signal can be 
mimicked here. The answer is no. The two-body decays of selectrons, 
relevant for us, are highly suppressed in this other scenario on account 
of the factor ${m_e}/{M_W}$ in the concerned couplings. The latter arises 
because $\tilde \chi^\pm_1$, $\tilde \chi^0_{1,2}$ are all almost 
exclusively higgsinos here. So selectrons primarily have three-body 
decays $\tilde e \rightarrow \nu_e W \tilde \chi^0_{1,2}$, $e Z \tilde 
\chi^0_{1,2}$ mediated by virtual heavier charginos/neutralinos 
$(\tilde \chi^\pm_2/\tilde \chi^0_2)$, which are gauginos, with finals 
states dominated by jets. One can easily estimate the
ratio of the partial widths of left selectron decays into two-body and
three-body channels to be  ${\cal O}(10^{-4})$ in this scenario
demonstrating that the desired two-body decays would be unobservable.
Therefore, our final state of a fast electron (muon) and a soft
pion distinguishes AMSB models from the light higgsino scenario. This
new result was highlighted in Ref. \cite{dkgprsr} and is more or less
true for the other signals discussed in this paper.  

We now come to the question of Standard Model background to our signal.
The signal can be classified into two categories. There is one in which
we see a heavily ionizing nearly straight charged track ending with a
soft pion with large impact parameter and $\mET$, the signal being
triggered with one or multiple fast electrons or muons. In the other case, 
while the other aspects remain the same, one may not see the heavily ionizing
charged track but the impact parameter of the soft pion can be resolved
and measured to be large. In the first case the heavily ionizing charged
track is due to the passage of a massive chargino with a very large
momentum. Due to this reason the charged tarck will be nearly straight in 
the presence of the magnetic field. One cannot imagine a similar situation 
in the SM with such a nearly straight heavily ionizing charged track due to 
a very massive particle. An ionized charged track can possibly arise from the
flight of a low energy charged pion, kaon or proton but it will curl
significantly in the magnetic field. Another distinguishing feature of
the charged track in our signal is that it will be terminated after a
few layers in the vertex detector and there will be a soft pion at the
end. In the second case, where the ionizing track is unseen, possible SM
backgrounds can come from the following processes: $e^+ + e^- \rightarrow
\tau^+ + \tau^-$ and $e^+ + e^- \rightarrow W^+ + W^-$. In the case of
$e^+ + e^- \rightarrow \tau^+ + \tau^-$, one $\tau$ can have the three
body decay $\tau \rightarrow e \nu_e \nu_\tau$ or $\mu \nu_\mu \nu_\tau$
and the other $\tau$ can go via the two body channel $\tau \rightarrow
\pi + \nu_\tau$. Thus we can have a final state of the type $e (\mu)
+ \pi + \mET$. Since we are considering an $(e^+ e^-)$ CM energy of 1
TeV, and the pion comes from a sequence of two-body production and decay, it
will have a fixed high momentum much in excess of 1 GeV. This will
clearly separate this type of background from our signal since in our case the
resulting pion is very soft with a momentum in the range of hundreds of
MeV. In the case of $e^+ + e^- \rightarrow W^+ + W^-$ a similar argument
follows. Here one $W$ can go to $e (\mu) + \nu_e (\nu_\mu)$ and the other 
one can go to $\tau + \nu_\tau$. The $\tau$ can subsequently go to one 
$\pi$ and a $\nu_\tau$, thereby producing the final state 
$e (\mu) + \pi + \mET$. As we have discussed just now, the resulting pion 
will have a very large momentum and again one can clearly separate the 
background from the signal.

\section{Conclusions}

In this paper we have presented a detailed study of possible signals
from the electroweak sector of the minimal AMSB model in a $1$ TeV CM
energy $e^+e^-$ linear collider. AMSB scenarios are attractive since
they do not have the FCNC problems of tree level gravity mediated
supersymmetry breaking models but retain their other virtues. One
interesting feature of most AMSB scenarios (including the minimal model)
is the occurance of two nearly degenarate winolike states, the LSP
neutralino $\N10$ and the lightest chargino $\C1pm$, as well as of the
long-lived decay $\C1pm \rightarrow \N10$ + soft $\pi^\pm$ resulting in
a displaced vertex $X_D$ with a heavy ionizing track and/or a detectable
soft $\pi^\pm$ with a distinctly large impact parameter. Each of our
signal events consists of fast leptons (any of which can be the trigger)
accompanied by $X_D$/soft $\pi$ numbering one or two.

Sleptons play a key role in our analysis. Sleptons in the minimal AMSB
model are predicted, on the basis of the required absence \cite{akundu}
of charge and color violating minima in the one-loop effective
potential, to be heavy and beyond the reach of a $500$ GeV CM energy
$e^+e^-$ collider. This is why we have considered $e^+e^-$ collision at
$\sqrt s$ = 1 TeV as in the proposed TESLA machine \cite{zerwas}. We
have calculated all relevant two-sparticle production cross sections and
have generated numbers for event rates at a given integrated luminosity
by considering all cascade decay modes. We have plotted sample
distributions for the transverse momentum of a final state lepton, that
of a soft $\pi$ as well as for the chargino decay length. Our generated
event numbers are large enough to enable us to make the following
definitive statement with confidence. An experimental effort along our
suggested directions will completely cover the remaining allowed region
of the parameter space of the minimal AMSB model.   


\section{Acknowledgement}

The work was initiated at the Sixth Workshop of High Energy Physics
Phenomenology (WHEPP-6) held at IMSc, Chennai, whose organisers are
thanked. AK acknowledges the hospitality of the Department of Theoretical 
Physics of the Tata Institute of Fundamental Research, Mumbai, where a part 
of this work was done. He was supported by the BRNS grant no.\ 2000/37/10/BRNS
of DAE, India. DKG acknowledges the hospitality of the Theory Group of KEK, 
JAPAN, where a part of this work was done. The work of DKG was supported in 
part by National Science Council under the grants NSC 89-2112-M-002-058 and 
in part by the Ministry of Education Academic Excellent Project 
89-N-FA01-1-4-3. Finally, we thank Sunanda Banerjee, Utpal Chattopadhyay, 
Kaoru Hagiwara, Gobinda Majumder and Pushan Majumdar for very helpful 
discussions.

\def\pr#1, #2 #3 { {\em Phys. Rev.}         {\bf #1},  #2 (19#3)}
\def\prd#1, #2 #3{ {\em Phys. Rev.}        {D \bf #1}, #2 (19#3)}
\def\pprd#1, #2 #3{ {\em Phys. Rev.}       {D \bf #1}, #2 (20#3)}
\def\prl#1, #2 #3{ {\em Phys. Rev. Lett.}   {\bf #1},  #2 (19#3)}
\def\pprl#1, #2 #3{ {\em Phys. Rev. Lett.}   {\bf #1},  #2 (20#3)}
\def\plb#1, #2 #3{ {\em Phys. Lett.}        {\bf B#1}, #2 (19#3)}
\def\pplb#1, #2 #3{ {\em Phys. Lett.}        {\bf B#1}, #2 (20#3)}
\def\npb#1, #2 #3{ {\em Nucl. Phys.}        {\bf B#1}, #2 (19#3)}
\def\pnpb#1, #2 #3{ {\em Nucl. Phys.}        {\bf B#1}, #2 (20#3)}
\def\prp#1, #2 #3{ {\em Phys. Rep.}        {\bf #1},  #2 (19#3)}
\def\zpc#1, #2 #3{ {\em Z. Phys.}           {\bf C#1}, #2 (19#3)}
\def\epj#1, #2 #3{ {\em Eur. Phys. J.}      {\bf C#1}, #2 (19#3)}
\def\mpl#1, #2 #3{ {\em Mod. Phys. Lett.}   {\bf A#1}, #2 (19#3)}
\def\ijmp#1, #2 #3{{\em Int. J. Mod. Phys.} {\bf A#1}, #2 (19#3)}
\def\ptp#1, #2 #3{ {\em Prog. Theor. Phys.} {\bf #1},  #2 (19#3)}
\def\jhep#1, #2 #3{ {\em J. High Energy Phys.} {\bf #1}, #2 (19#3)}
\def\pjhep#1, #2 #3{ {\em J. High Energy Phys.} {\bf #1}, #2 (20#3)}


\end{document}